\documentclass[12pt,a4paper]{article}
\usepackage{amssymb,amscd}

\title{
Non-autonomous H\'{e}non-Heiles Systems}

\author{Andrew N. W. Hone\\
\normalsize
\em Department of Mathematics and Statistics,\\
\normalsize
\em The University of Edinburgh, \\
\normalsize
\em Edinburgh, UK. \\
\normalsize
e-mail: hone@maths.ed.ac.uk
\\
}

\begin{document}
\renewcommand{\theequation}{\arabic{section}.\arabic{equation}}
\renewcommand{\thepage}{}
\begin{titlepage}
\maketitle

\begin{abstract}

Scaling 
similarity solutions of three integrable    
PDEs, namely the Sawada-Kotera, fifth order KdV and Kaup-Kupershmidt 
equations, are considered.  
It is shown that the resulting ODEs may be written as  
non-autonomous Hamiltonian equations, which are 
time-dependent generalizations  
of the well-known integrable H\'{e}non-Heiles systems.  
The (time-dependent) Hamiltonians  
are given by logarithmic    
derivatives of the tau-functions (inherited  
from the original PDEs).  
The ODEs for the similarity solutions also have inherited 
B\"{a}cklund transformations, which may be used 
to generate sequences of 
rational solutions as well as other special solutions  
related to the first Painlev\'{e} transcendent.    

\end{abstract}
\vfill
\end{titlepage}
\renewcommand{\thepage}{\arabic{page}}

\section{Introduction}

The six Painlev\'{e} transcendents (PI-VI)    
have had a considerable amount of attention 
devoted to them in recent years, and have been studied from many different 
points of view. Their original discovery came about 
from Painlev\'{e}'s classification of second order ODEs   
(of a particular form) having no movable 
critical points. They have also been approached by way of 
isomonodromic deformation of linear differential   
equations \cite{fla}, or 
via abelian integrals and algebraic geometry \cite{manin}. 
Furthermore, they have found numerous physical applications.  
For example, the first Painlev\'{e} transcendent (PI),  
\begin{equation}    
\frac{d^{2}w}{dz^{2}}=6w^{2}+z, \label{eq:peeone}    
\end{equation}    
arises as a string equation   
in the matrix models of 2-D gravity \cite{douglas},    
while other Painlev\'{e} transcendents govern the   
behaviour of correlation functions in certain models  
of statistical mechanics \cite{babe}, quantum   
field theory \cite{IIKS} and random matrices \cite{harnad} (see chapter 7 
of \cite{abcla} for a more complete survey).   
Higher order ODEs of Painlev\'{e} type 
also occur in many situations,  
but as yet there is no complete classification of such equations.

Another context in which ODEs of Painlev\'{e} type arise naturally 
is as similarity reductions of integrable PDEs. 
A great deal of the renewed interest in  
Painlev\'{e} equations was generated by the conjecture of   
Ablowitz, Ramani and Segur
\cite{ars},
which states that all similarity reductions of integrable PDEs are of  
Painlev\'{e} type.  
For example, the modified Korteweg-deVries (mKdV) equation,
\begin{equation}
v_{t}=v_{xxx}-6v^{2}v_{x},   \label{eq:mk}
\end{equation}
has scaling similarity solutions of the form
\[ 
v(x,t)=(-3t)^{-\frac{1}{3}}y(z(x,t)), \qquad z=(-3t)^{-\frac{1}{3}}x,   
\] 
where $y(z)$ satisfies  
\begin{equation}
y''=2y^{3}+zy+\alpha,   \label{eq:PII}
\end{equation}
for some constant $\alpha$
($'$ denotes $\frac{d}{dz}$),
which is the second Painlev\'{e}
equation (PII).
An important feature of the   
similarity reductions of soliton-type equations is that they  
inherit much of the stucture associated with integrability, such as 
B\"{a}cklund transformations and solutions in terms of special functions. 

In the work of Okamoto \cite{oka},
each of the equations PI-VI
is written as a non-autonomous
Hamiltonian system, with a polynomial Hamiltonian, e.g.    
(\ref{eq:PII}) may be obtained from the 
Hamiltonian  
\begin{equation}   
h=\frac{1}{2}p^{2}-(y^{2}+\frac{z}{2})p-(\alpha+\frac{1}{2})y  
\label{eq:PIIham} 
\end{equation}   
(where $y$, $p$ are the canonically conjugate variables, and  
$z$ denotes time).  
Each of the Hamiltonians used in  
\cite{oka} has explicit dependence 
on the time $z$, and is obtained from a   
holomorphic tau-function,  
\[ 
h(z) =\frac{d}{dz} \log[\tau(z)]. 
\] 
The tau-function provides a simple    
way to encode both the solutions of the equations and their  
B\"{a}cklund transformations.   
In the case of PII, this tau-function essentially coincides with the 
tau-function of mKdV/KdV (up to a factor of  
$\exp[z^{3}/24]$).     
Manin \cite{manin} has found an alternative  
Hamiltonian description for PVI, involving the Weierstrass 
$\wp$ function, but  
his approach is difficult  
to generalize to higher order  
equations of Painlev\'{e} type.   

In what follows we   
apply the approach  
of Okamoto, using polynomial Hamiltonians and tau-functions,   
in order to  
analyze the 
the scaling
similarity solutions of
three different integrable, fifth order
evolution equations, namely
the Sawada-Kotera (SK) equation,   
\begin{equation}
u_{t}=u_{5x}+5uu_{3x}+5u_{x}u_{xx}+5u^{2}u_{x}, \label{eq:sawkot}
\end{equation}
fifth order KdV,  
\begin{equation}
u_{t}=u_{5x}+10uu_{3x}+20u_{x}u_{xx}+30u^{2}u_{x}, \label{eq:thefivk}
\end{equation}
and the  
Kaup-Kupershmidt (KK) equation,  
\begin{equation}
u_{t}=u_{5x}+10uu_{3x}+25u_{x}u_{xx}+20u^{2}u_{x}. \label{eq:kauku}
\end{equation} 
The ODEs for scaling similarity solutions of
each of the flows in
the SK, KdV and KK hierarchies, as well as their  
modified hierarchies (related by a Miura map),    
are presented in section 2.
In the third section, by a direct extension of  
Fordy's results \cite{fordy1} on stationary flows, it is shown that   
the scaling similarity solutions of the fifth order equations 
in each hierarchy are related to non-autonomous  
generalizations of the integrable H\'{e}non-Heiles systems.   
We describe the tau-functions, Miura maps  
and B\"{a}cklund transformations   
associated to these systems, and use this  
machinery to generate sequences of special solutions   
in section 4.  
The essence of our method is that the Hamiltonian  
formalism for stationary flows \cite{fordy2} may be extended to  
other (non-stationary) similarity reductions.  
We  briefly       
discuss further developments in our concluding section.     

\vspace{.1in} 
{\bf Note.} In their most basic form, these results were  
originally presented in \cite{hone}, although these  
proceedings do not seem to be generally available at present;    
my thesis \cite{honeth} contains a more thorough discussion.  
Shortly after my thesis was submitted, I was made aware of Kudryashov's  
paper \cite{kudry} concerning similarity reductions in the KdV/mKdV  
hierarchy. The results in \cite{kudry} have some overlap  
with my own, but do not make use of the Hamiltonian formalism.     
        
\section{Scaling Similarity Solutions in the Sawada-Kotera,  
KdV and Kaup-Kupershmidt
Hierarchies}
\setcounter{equation}{0}  

In this section 
we develop a concise 
notation to describe the scaling similarity 
solutions of any one of the flows of the SK, KdV or KK hierarchies. 
We then apply it to some 
particular examples, including the similarity 
solutions of the fifth order equations which we 
study in detail in sections 3 and 4.          
    
\subsection{General Description of Scaling Similarity Solutions} 

SK and KK have only one Hamiltonian structure, but KdV is
bi-Hamiltonian,
and here we will be using the second Hamiltonian structure. 
Following Fordy \cite{fordy1}, we are able to consider all three 
hierarchies at once.  
The $n$th flow in each of the hierarchies can be written as
\begin{equation}
\frac{\partial u}{\partial t_{n}} = (\partial_{x}^{3} + 8au\partial_{x} +
4au_{x})\delta_{u}H_{n}[u],            \label{eq:ham}
\end{equation}
where $a=1/2$ for SK and KdV, $a=1/4$ for KK, and $H_{n}$ is the $n$th
Hamiltonian density   
for the hierarchy in question.
There is also a Miura map from the modified versions of the hierarchies,
given by
\[
u=-v_{x}-2av^{2}=:M[v].
\]
The $n$th
flow of the modified hierarchy is  
\begin{equation}
\frac{\partial v}{\partial t_{n}}=(-\partial_{x})\delta_{v}H_{n}[M[v]].
\label{eq:mham}
\end{equation}
The Miura map means that given $v$ satisfying (\ref{eq:mham}) for each $n$,
the corresponding $u=-v_{x}-2av^{2}$ satisfies (\ref{eq:ham}).
Further details on these hierarchies may be found in \cite{FG}.    

The $n$th flow of the hierarchy is unchanged by the scaling
\[
x\rightarrow \beta x, \qquad  
t_{n}\rightarrow \beta^{m}t_{n}, \qquad  
u\rightarrow \beta^{-2} u,
\]
where
$m=m(n)$ is a scale weight dependent on the hierarchy and on 
which flow is being considered.  
Similarly the modified
flow is invariant under the same scaling but with
\[
v\rightarrow\beta^{-1} v.
\]
Hence there are scaling similarity solutions of     
the form 
\[
u(x,t_{n})=\theta^{2}(t_{n})w(z), \qquad z=x\theta(t_{n}),  
\qquad \frac{d\theta}{dt_{n}}=\theta^{m+1}.
\]
(We use the notation of \cite{clark}.) The
corresponding
similarity solution for the modified flow is
\[
v=\theta(t_{n})y(z),
\]
with
the scaled Miura map giving 
\[ 
w=-y'-2ay^{2}=:\tilde{M}[y]   
\] 
($'$ denotes $\frac{d}{dz}$
throughout). 

In the context of an integrable hierarchy, it is 
customary to think of the dependent variable as a function of all the times, 
\[ 
u=u(t_{1}=x,t_{2},t_{3},...). 
\]     
For the scaling similarity solutions of the $n$th flow, 
it is necessary to drop dependence   
on anything other than $x$ and $t_{n}$, 
and it is inconsistent to consider the other flows simultaneously.   
To incorporate the other 
flows extra similarity variables must be included     
(e.g. in similarity reductions of the mKP 
equation \cite{kudry}). Such multi-phase   
similarity reductions are considered in  
chapter 4 of \cite{newell}, and we will discuss how our approach  
extends to these solutions in \cite{hone3}.   
Here we consider dependence on $x$ and $t_{n}$ only,  
and henceforth we drop the suffix $n$, bearing in mind that the 
actual form of $H$ depends on which particular flow we have 
chosen. 

Substituting the similarity forms into
the equations of motion (\ref{eq:ham}) and (\ref{eq:mham})
(and cancelling out powers of $\theta$ on either side)
yields the ODEs for $w$ and $y$. If we let $\tilde{H}$   
denote the rescaled $H$  
(in terms of $w$ with 
powers of $\theta$ divided out) then
we obtain the following:  
\begin{eqnarray}
(\partial^{3}+8aw\partial+4aw')\left(\delta_{w}\tilde{H}[w]   
-\frac{1}{4a}z\right) & = & 0,  
\label{eq:sim} \\  
\partial(\delta_{y}\tilde{H}[\tilde{M}[y]]+zy) & = & 0.    \label{eq:msim}
\end{eqnarray}
The symbol $\partial$ denotes derivatives with respect to $z$. 
Both equations (\ref{eq:sim},\ref{eq:msim}) 
can be integrated once, to yield  
\begin{eqnarray}
\frac{d^{2}f}{dz^{2}}+4awf+\frac{\lambda^{2}-(\frac{df}{dz})^{2}}{2f}  
 & = & 0,   
\label{eq:intsim} \\   
\frac{df}{dz}-4ayf+\lambda & = & 0.    \label{eq:mintsim}
\end{eqnarray}
In the above, $\lambda$ is a constant of integration,  
and we have introduced   
the quantity   
\begin{equation}  
f:=\delta_{w}\tilde{H}-\frac{1}{4a}z. \label{eq:theffun}  
\end{equation}  

The equations (\ref{eq:sim}, \ref{eq:msim}) deserve some comment, as  
we shall make use of them in section 3.  
With the definition (\ref{eq:theffun}), we may take     
$f=f[w]$,  
and thus (\ref{eq:intsim}) is to be considered as  
the ODE for $w$, obtained directly from   
(\ref{eq:sim}) after an integration.    
However, using the scaled Miura map    
it is also consistent to take $f=f[\tilde{M}[y]]$, and so   
(by a slight abuse of notation) the equation  
(\ref{eq:mintsim}) represents the ODE for the variable $y$,    
obtained by integrating (\ref{eq:msim}). However,   
a more obvious direct integration of (\ref{eq:msim})
would be
\begin{equation}
\delta_{y}\tilde{H}+zy+\alpha=0,
\label{eq:altern}
\end{equation}
for some constant $\alpha$. To show that  
(\ref{eq:mintsim}) is equivalent to (\ref{eq:altern}),     
it is necessary to use the scaled version of the   
identity  
\[    
\delta_{v}H=(M')^{*}\delta_{u}H=(\partial_{x}-4av)\delta_{u}H  
\]   
(with $M'$ being the Fr\'{e}chet derivative of $M$), and then it is  
found that the constants of integration are related by   
\[ 
\alpha=\lambda-\frac{1}{4a}. 
\] 

Another important point is that the same constant
$\lambda$ appears in both (\ref{eq:intsim}) and (\ref{eq:mintsim}).    
This is because the Miura map for the original  
PDEs (\ref{eq:ham},\ref{eq:mham}) becomes a one-one
correspondence 
between the ODEs for their scaling similarity solutions.    
The scaled Miura map is   
\begin{equation} 
w=-y'-2ay^{2}, \label{eq:thscmi}  
\end{equation}   
and it has an inverse given by 
\begin{equation}   
y=\frac{f'+\lambda}{4af}.  
\label{eq:thinvm} 
\end{equation} 
In  (\ref{eq:thinvm}) $f$ should be regarded as  $f[w]$,  
and it is necessary to assume 
$f\not\equiv 0$ since otherwise this equation breaks down.    
It is straightforward to see the one-one  
correspondence between  
(\ref{eq:intsim}) and (\ref{eq:mintsim}).   
Indeed, given a solution $w$ of (\ref{eq:intsim}),                   
the modified variable $y$ may be defined by the inverse Miura map  
(\ref{eq:thinvm}).   
A direct calculation then shows that the Miura formula  
(\ref{eq:thscmi}) holds, meaning that $f$ 
can be reinterpreted as $f[\tilde{M}[y]]$. Thus (on rearranging  
(\ref{eq:thinvm})) $y$ satisfies the ODE (\ref{eq:mintsim}); the  
converse follows by reversing this argument (or by scaling the  
Miura map for the PDEs).       
To make things more concrete, it is worth looking at some particular cases.

\subsection{Similarity Solutions of KdV and PII} 

The correspondence between PII (the equation (\ref{eq:PII}))    
and scaling similarity solutions of the
ordinary KdV equation is well-known.   
Taking $H=\frac{1}{2}u^{2}$   
in (\ref{eq:ham}) with $a=1/2$ yields KdV,
\begin{equation}       
u_{t}=u_{xxx}+6uu_{x}, \label{eq:kdvagain}     
\end{equation}  
with scaling similarity solutions   
\[
u(x,t)=(-3t)^{-\frac{2}{3}}w(z), \, \, z=(-3t)^{-\frac{1}{3}}x.   
\]
The scaled Hamiltonian $\tilde{H}=\frac{1}{2}w^{2}$  gives   
\[
f[w]=w-\frac{z}{2},
\]
whence (\ref{eq:intsim}) with $a=\frac{1}{2}$ becomes 
\begin{equation}
w''+2w^{2}-zw+\frac{\alpha(\alpha+1)+w'-(w')^{2}}{2w-z}=0,
\label{eq:ks}
\end{equation}
taking $\lambda=\alpha+\frac{1}{2}$.

For the scaling
similarity solutions  of mKdV (\ref{eq:mk}),  
the one-one correspondence
between solutions of PII     
and (\ref{eq:ks}) is given by
\[
w=-y'-y^{2}
\]
(the scaled Miura map) and
\[
y=\frac{w'+\alpha}{2w-z},  
\]
the latter being a particular case 
of the inverse Miura 
formula (\ref{eq:thinvm}). 
Observe that, in terms of $y$, we have
\[
f[\tilde{M}[y]]=-y'-y^{2}-\frac{z}{2},
\]
and on putting this into (\ref{eq:mintsim}) with $a=\frac{1}{2}$, 
PII results.
This example is also considered in \cite{abfo}, for instance.

\subsection{Similarity Solutions of Fifth Order Equations} 

Our second example constitutes the main subject of sections 3 and 4.  
Following \cite{fordy1}, we use the equation  
(\ref{eq:ham}) with Hamiltonian density  
\[
H=-\frac{1}{2}u_{x}^{2}-\frac{1}{3}bu^{3},
\]
leading to    
the fifth order equation    
\begin{equation}
u_{t}=(u_{xxxx}+(8a-2b)uu_{xx}-2(a+b)u_{x}^{2}-\frac{20}{3}abu^{3})_{x}.
\label{eq:fif}
\end{equation}
There are only three cases (i),(ii),(iii), corresponding to 
\begin{equation}     
a  =  \frac{1}{2}, \quad  
\frac{1}{2}, \quad \frac{1}{4},  
\qquad b  =  -\frac{1}{2}, \quad -3, \quad -4,   
\label{eq:choices}   
\end{equation}      
for which an equation of the form
(\ref{eq:fif}) is integrable, 
and these three cases correspond to the   
SK equation (\ref{eq:sawkot}),     
the fifth order KdV equation (\ref{eq:thefivk}),    
and the KK equation (\ref{eq:kauku})    
respectively.   
When specializing to one of the three hierarchies, it is necessary 
to take the correct values of
$a$ and $b$ in each case. While some of the calculations we 
present are valid 
for arbitrary $a$, $b$, all properties relevant to the 
integrability of the equations are lost in general. 
  
The scaling similarity solutions of (\ref{eq:fif}) take the form  
\[
u(x,t)=(-5t)^{-\frac{2}{5}}w(z), \, \, z=(-5t)^{-\frac{1}{5}}x.   
\]
From the scaled density $\tilde{H}$ we find     
\[
f[w]=w''-bw^{2}-\frac{z}{4a},
\]
and substituting this into (\ref{eq:intsim}) yields   
a fourth order ODE for $w$, which we prefer not to 
present explicitly. In the next section we show that this    
ODE for $w$ is equivalent to a non-autonomous Hamiltonian system   
of H\'{e}non-Heiles type.     

Scaling similarity solutions of 
the associated modified equations take the form   
\[ 
v(x,t)=(-5t)^{-\frac{1}{5}}y(z).
\]
Using the scaled Miura map, we may express $f$ 
in terms of the modified variable $y$:
\begin{equation}  
f[\tilde{M}[y]]=-y'''-4ayy''-(4a+b)(y')^{2}-4aby^{2}y'   
-4a^{2}by^{4}-\frac{z}{4a}. 
\label{eq:thefydef}  
\end{equation}  
Then the ODE for $y$ is (from (\ref{eq:mintsim}))
\begin{equation}
y^{(iv)}=-2(6a+b)y'y''+4a(4a-b)(y^{2}y''+y(y')^{2})+16a^{3}by^{5}+zy+\alpha,
\label{eq:fy}
\end{equation}
with 
\[ 
\alpha=\lambda-\frac{1}{4a}. 
\] 
Given a solution to (\ref{eq:fy}) we can
then obtain a solution to the fourth order ODE for $w$,  
via $w=-y'-2ay^{2}$ (and vice-versa, using the inverse Miura map   
(\ref{eq:thinvm})).    

On substituting in the relevant values of $a$ and $b$   
for the integrable cases
(i),(ii),(iii)  into the equation 
(\ref{eq:fy}) we find  
\begin{eqnarray}
y^{(iv)} & = & -5y'y''+5(y^{2}y''+y(y')^{2})-y^{5}+zy+\alpha, \label{eq:mSK} 
\\ 
y^{(iv)} & = & 10(y^{2}y''+y(y')^{2})-6y^{5}+zy+\alpha, \label{eq:mKdV} \\  
y^{(iv)} & = & 5y'y''+5(y^{2}y''+y(y')^{2})-y^{5}+zy+\alpha, \label{eq:mKK}   
\end{eqnarray} 
the ODEs for scaling similarity solutions of  
the modified SK, fifth order mKdV and  
modified KK equations respectively.  
Notice that (\ref{eq:mSK}) and (\ref{eq:mKK}) differ only
by a sign in the even ($y'y''$)
terms. Hence if $y_{(i)}$ is as solution to (\ref{eq:fy}) for case (i), then 
$y_{(iii)}=-y_{(i)}$ will be a solution to that equation for case (iii) with 
$\alpha$ replaced by $-\alpha$.   
This is because the modified
hierarchies in these two cases are essentially the same; both 
SK and KK have a third order Lax operator, which factorizes to yield the 
Miura map (see \cite{FG}).   
In the next section   
we shall use this connection between case (i) and case (iii)  
to derive the B\"{a}cklund
transformations for these scaling similarity equations,   
after relating them to H\'{e}non-Heiles systems.   

\vspace{.1in} 
{\bf Remark.} In \cite{kudry} the sequence of equations  
(\ref{eq:mintsim}) are referred to as the second Painlev\'{e}  
equations of $2n$th order (corresponding to scaling similarity  
solutions to the flow given by $H_{n+1}$). For instance,  
(up to a rescaling) the equation (\ref{eq:mKdV}) is   
the second Painlev\'{e} equation of fourth order in the   
nomenclature of \cite{kudry}. However, this sequence of ODEs is by no means  
new; it appears in \cite{fla} under the name  of the Painlev\'{e} II 
Family.   
          
\section{H\'{e}non-Heiles Systems}
\setcounter{equation}{0}  

\subsection{Stationary Flows and Integrable H\'{e}non-Heiles} 
 
The original H\'{e}non-Heiles system is given by a
Hamiltonian with two degrees of
freedom:

\begin{equation}
h=\frac{1}{2}(p_{1}^2+p_{2}^{2}) + aq_{1}q_{2}^{2} -\frac{1}{3}bq_{1}^{3},
\label{eq:henon}
\end{equation}
The equations of motion are just Hamilton's equations,   
\begin{equation}
\frac{dq_{j}}{dz}  =  \frac{\partial h}{\partial p_{j}}, \qquad  
\frac{dp_{j}}{dz}  =  -\frac{\partial h}{\partial q_{j}}. \label{eq:hameq1}  
\end{equation}
(We are denoting the time by $z$ here to make connection with our other 
results.) 
It has been known for some time from Painlev\'{e} analysis \cite{cha}
that this system
is integrable for only three values of the ratio $r=a/b$,   
namely  
\[ 
r=-1, \quad -1/6, \quad -1/16. 
\]  
More recently it was shown by  Fordy \cite{fordy1}   
that in these integrable
cases the equations of motion are just disguised 
versions of the stationary flows  
of the SK, fifth order
KdV and KK equations.   
Thus the special  
values of $a$ and $b$ listed   
in the previous section give the right values for
the ratio $r$ in the cases (i),(ii),(iii) respectively.
The zero curvature form of these PDEs yields a matrix Lax representation
of the stationary flows, and then traces of powers of the Lax matrix give the
Hamiltonian and a second constant of motion   
in involution   
(implying Liouville integrability).
It was subsequently shown that all three Hamiltonian   
systems are completely separable in suitable
coordinates, and may be integrated in terms of theta functions of genus one
(cases (i) \& (iii)) or genus two (case (ii))\cite{caboz}.

\subsection{Non-autonomous Hamiltonians for 
Scaling Similarity Solutions} 

In order to consider the   
ODEs for scaling similarity solutions of the three 
integrable fifth order PDEs (all of the form (\ref{eq:fif})),
we adapt Fordy's approach \cite{fordy1}  
to the stationary flows,   
and write them in Hamiltonian form.
Recall that the previous section   
we have the quantity $f$ given by  
\begin{equation}   
f=w''-bw^{2}-\frac{z}{4a}, \label{eq:fdef}  
\end{equation}   
and on substituting this into 
(\ref{eq:intsim})  
a full fourth order ODE for $w$ results.   
Instead of presenting this  
explicitly,    
we set 
\[ 
w=q_{1}, \qquad f=-aq_{2}^{2};    
\]
the latter is a squared eigenfunction substitution \cite{baker2}.  
Then we may rewrite (\ref{eq:intsim},\ref{eq:fdef}) as a coupled system for 
$q_{1},q_{2}$:   
\begin{eqnarray}  
q_{1}'' & = & bq_{1}^{2}-aq_{2}^{2}+\frac{z}{4a}, \label{eq:syeq1} \\   
q_{2}'' & = & -2aq_{1}q_{2}-\frac{\lambda^{2}}{4a^{2}q_{2}^{3}}.
\label{eq:syeq2}
\end{eqnarray}
This coupled system just follows from Hamilton's equations 
(\ref{eq:hameq1}), where now the 
Hamiltonian is given by  
\begin{equation}  
h=\frac{1}{2}(p_{1}^{2}+p_{2}^{2})+aq_{1}q_{2}^{2}-\frac{1}{3}bq_{1}^{3}
-\frac{\lambda^{2}}{8a^{2}}q_{2}^{-2}-\frac{1}{4a}zq_{1}. 
\label{eq:nahh}  
\end{equation}  
Compared with (\ref{eq:henon}), this has an extra inverse square term and a
non-autonomous (time-dependent) term in the potential. 

Thus we have shown that the equations for the similarity
solutions of (\ref{eq:fif})   
may be viewed as
non-autonomous H\'{e}non-Heiles systems. Because of the explicit
time-dependence, the Hamiltonian $h(z)$  
satisfies a fourth order ODE which (in the generic  
situation $f'\not\equiv 0$) is   
equivalent to the system (\ref{eq:syeq1},\ref{eq:syeq2}) (this is  
proved in \cite{honeth}).    
Applying Painlev\'{e} analysis to this system (which is  
almost identical to the analysis for    
the autonomous case \cite{cha}),    
it is found that   
the Painlev\'{e} test is only satisfied   
for the three special values of the ratio $r$ (which correspond to the  
scaling similarity solutions of the integrable PDEs).     
Because they are similarity reductions of   
PDEs solvable by inverse scattering,  
these special cases can also be derived from a  
zero curvature condition \cite{honeth},    
and solved by the inverse monodromy method \cite{fla}, but we  
will not   
consider this approach here.   
           
For the particular special choices    
of $a$ and $b$ given in (\ref{eq:choices}),    
we denote the Hamiltonian systems      
(with Hamiltonian (\ref{eq:nahh})) by 
$\mathcal{H}\mathnormal{_{(i)}}$,
$\mathcal{H}\mathnormal{_{(ii)}}$, 
$\mathcal{H}\mathnormal{_{(iii)}}$ (corresponding to  
the reductions of SK, KdV and KK respectively).   
Each system has B\"{a}cklund transformations    
which can be viewed as a canonical 
transformations in the variables $q_{j},p_{j},z$,    
i.e. under the transformation    
\[ 
(q_{j},p_{j},z)\longrightarrow(\tilde{q}_{j},\tilde{p}_{j},\tilde{z}), 
\]       
the two-form 
\begin{equation}    
\omega=\sum_{j=1,2}dp_{j}\wedge dq_{j}-dh\wedge dz 
\label{eq:twoform}  
\end{equation}   
is preserved.   

The construction of the B\"{a}cklund transformations is  
possible due to the existence of the Miura maps.  
For the system $\mathcal{H}\mathnormal{_{(ii)}}$     
we have (by the general arguments of section 2)  
a one-one correspondence between its solutions and the solutions of the 
modified similarity equation (\ref{eq:mKdV}), while 
both $\mathcal{H}\mathnormal{_{(i)}}$ 
and $\mathcal{H}\mathnormal{_{(iii)}}$ have similar correspondences with 
(\ref{eq:mSK}).  
We are also able to 
define a tau-function for the Hamiltonians,  via 
\[ 
h(q_{j}(z),p_{j}(z),z)\propto(\log[\tau(z)])'   
\] 
(for some constant of proportionality,  
dependent on which of the three cases is being 
considered). We consider the system  
$\mathcal{H}\mathnormal{_{(ii)}}$ first, as this is the simplest,   
while the systems $\mathcal{H}\mathnormal{_{(i),(iii)}}$ are then  
considered in tandem (because they are connected via the correspondence  
with the same modified equation (\ref{eq:mSK})).

\subsection{The Hamiltonian System $\mathcal{H}\mathnormal{_{(ii)}}$} 
 
The Hamiltonian for $\mathcal{H}\mathnormal{_{(ii)}}$ is given 
explicitly by 
\begin{equation}   
h_{\lambda}=\frac{1}{2}(p_{1}^{2}+p_{2}^{2})+\frac{1}{2}q_{1}q_{2}^{2}
+q_{1}^{3}
-\frac{\lambda^{2}}{2}q_{2}^{-2}-\frac{1}{2}zq_{1}.  
\label{eq:hamKdV} 
\end{equation}  
We will henceforth put suffixes on all quantities to denote their 
dependence on $\lambda$, as the B\"{a}cklund transformation 
relates the same quantities for different values of this 
parameter. There is also the alternative parameter 
$\alpha=\lambda-\frac{1}{2}$ appearing in (\ref{eq:mKdV}), but 
$\lambda$ is the more natural one in that  
(\ref{eq:hamKdV}) depends only on $\lambda^{2}$, i.e. 
\[ 
h_{\lambda}=h_{-\lambda},  
\] 
and thus the same is true for the solutions $(q_{j}(z),p_{j}(z))$ to the 
system with Hamiltonian $h_{\lambda}$.   
The most convenient variables to use for this system are 
\[ 
w_{\lambda}=q_{1}, \qquad f_{\lambda}=-\frac{1}{2}q_{2}^{2},  
\]     
where (by (\ref{eq:fdef}) with $a=1/2$, $b=-3$) 
we have 
\begin{equation}   
f_{\lambda}=w_{\lambda}''+3w_{\lambda}^{2}-\frac{z}{2}. 
\label{eq:kfdef}   
\end{equation}   

Using the Miura map, there is a one-one correspondence 
with solutions to (\ref{eq:mKdV}). 
In other words, given a solution to the 
Hamiltonian system, we find a solution $y=y_{\lambda}$ to equation 
(\ref{eq:mKdV}) (with $\alpha=\lambda-\frac{1}{2}$) via the 
formula (\ref{eq:thinvm}),  
\begin{equation}   
y_{\lambda}=\frac{f_{\lambda}'+\lambda}{2f_{\lambda}}.  
\label{eq:invmiura}  
\end{equation}  
We also have the usual 
Miura expression, 
\begin{equation}   
w_{\lambda}=-y_{\lambda}'-y_{\lambda}^{2}, \label{eq:mmap}  
\end{equation}   
which means that conversely a solution to (\ref{eq:mKdV}) determines 
a solution to the system $\mathcal{H}\mathnormal{_{(ii)}}$.          
Before presenting the B\"{a}cklund transformation, it is helpful to 
introduce the tau-function. 

\newtheorem{defn}{Definition}   

\begin{defn} 
 For a solution $(q_{j}(z),p_{j}(z))$  
of the system $\mathcal{H}\mathnormal{_{(ii)}}$, we have 
the Hamiltonian $h_{\lambda}(z)=h_{\lambda}(q_{j}(z),p_{j}(z),z)$. 
The tau-function associated with this solution is given by 
\[ 
h_{\lambda}(z)=-\frac{d}{dz}\log[\tau_{\lambda}(z)]. 
\] 
\end{defn} 

The above definition is chosen to be consistent with the tau-function 
of the KdV hierarchy, where the dependent variable $u$ is 
expressed as 
\begin{equation} 
u(x,t)=2(\log[\tau(x,t)])_{xx}. \label{eq:KdVtau} 
\end{equation} 
For scaling 
similarity solutions we require that $\tau$ depends on $x$ and $t$ 
only through the combination $z=x\theta(t)$, which means we must have   
\[ 
w_{\lambda}=2(\log[\tau(z)])''.   
\] 
This agrees with our definition, for  
differentiating the Hamiltonian gives  
\[ 
-(\log[\tau(z)])''=h_{\lambda}'=-\frac{1}{2}q_{1}=-\frac{1}{2}w_{\lambda}. 
\] 
 
\newtheorem{propn}{Proposition} 
\begin{propn} 
 The B\"{a}cklund transformation for the equation 
(\ref{eq:mKdV}) may be written in the form     
\begin{equation}   
y_{\lambda+1}=-y_{\lambda}+\frac{\lambda}{f_{\lambda}}. 
\label{eq:mKdVbackl}  
\end{equation}   
Moreover, this induces a canonical transformation from the  
system $\mathcal{H}\mathnormal{_{(ii)}}$ with parameter $\lambda$ 
to the same system with parameter $\lambda+1$:   
\[ 
q_{j}\rightarrow \tilde{q}_{j}, \quad 
p_{j}\rightarrow \tilde{p}_{j}, \quad
z\rightarrow z, 
\quad
h_{\lambda}\rightarrow h_{\lambda+1}. 
\] 
The modified variable $y_{\lambda}$ may also be written in terms of 
the two tau-functions related by this transformation: 
\begin{equation}   
y_{\lambda}=(\log[\tau_{\lambda-1}/\tau_{\lambda}])'. 
\label{eq:mKdVytau} 
\end{equation}   
\end{propn} 

\noindent  
{\bf Proof.}   
The first thing to observe is that $w_{\lambda}$ is related to two 
different modified variables by the Miura map (\ref{eq:mmap}): 
\[ 
w_{\lambda}=-y_{\lambda}'-y_{\lambda}^{2} 
=-y_{-\lambda}'-y_{-\lambda}^{2}. 
\] 
This is a well-known property of Miura maps, but can be seen directly 
from the fact that the fourth order ODE for $w_{\lambda}$ 
just depends on $\lambda^{2}$, so 
\[ 
w_{\lambda}=w_{-\lambda}. 
\] 
The inverse Miura map (\ref{eq:invmiura}) gives 
\begin{equation}   
y_{\pm\lambda}=\frac{f_{\lambda}'\pm\lambda}{2f_{\lambda}}, 
\label{eq:taupm}     
\end{equation}   
the solutions to (\ref{eq:mKdV}) for parameter 
$\alpha=\pm\lambda-\frac{1}{2}$. Looking at the modified 
equation (\ref{eq:mKdV}), we see that it is unchanged on sending 
\[ 
y\rightarrow-y, \qquad \alpha\rightarrow-\alpha. 
\] 
Hence $-y_{-\lambda}$ will be a solution to this 
equation for $\alpha=\lambda+\frac{1}{2}=(\lambda+1)-\frac{1}{2}$, 
or in other words we may identify   
\[ 
y_{\lambda+1}=-y_{-\lambda}.  
\] 
It is then straightforward to derive (\ref{eq:mKdVbackl}) using 
the inverse Miura formula. This same argument also works for 
PII and the rest of the PII Family. Note that   
$f_{\lambda}$ in the right hand side of 
the B\"{a}cklund transformation may be written in terms of 
$y_{\lambda}$ and its derivatives. 

To define the induced transformation of 
the Hamiltonian system $\mathcal{H}\mathnormal{_{(ii)}}$,  
first of all it is necessary to write $y_{\lambda+1}$ in terms of 
the variables appearing in $h_{\lambda}$: 
\begin{equation}      
y_{\lambda+1}  = -\frac{p_{2}q_{2}+\lambda}{q_{2}^{2}}.  
\label{eq:yfuncpq} 
\end{equation}   
The Miura map produces a new solution to the fourth order equation 
(\ref{eq:intsim}) at parameter value $\lambda+1$, 
\[ 
w_{\lambda+1}=-y_{\lambda+1}'-y_{\lambda+1}^{2}. 
\]    
Now we can define the new variables $(\tilde{q}_{j},\tilde{p}_{j})$     
via 
\[ 
\tilde{q}_{1}=w_{\lambda+1}, \quad \tilde{p}_{1}=w_{\lambda+1}', 
\quad \tilde{q}_{2}=(-2f_{\lambda+1})^{\frac{1}{2}}, \quad 
\tilde{p}_{2}=-(-2f_{\lambda+1})^{-\frac{1}{2}}f_{\lambda+1}',     
\] 
and these will clearly satisfy the system $\mathcal{H}\mathnormal{_{(ii)}}$ 
with Hamiltonian $h_{\lambda+1}$. It is simple to write the new 
variables in terms of the old (and vice-versa),  
and we present the formulae here for 
completeness: 
\begin{equation}  
\begin{array}{llllll}   
\tilde{q}_{1} & = & -q_{1}-2y_{\lambda+1}^{2}, &   
q_{1} & = & -\tilde{q}_{1}-2y_{\lambda+1}^{2},   \\ 
\tilde{p}_{1} & = & -p_{1}-4y_{\lambda+1}(q_{1}+y_{\lambda+1}^{2}), &  
p_{1} & = & -\tilde{p}_{1}+4y_{\lambda+1}(\tilde{q}_{1}+y_{\lambda+1}^{2}), 
 \\ 
\tilde{q}_{2} & = & \Upsilon^{\frac{1}{2}}, &  
q_{2} & = & \tilde{\Upsilon}^{\frac{1}{2}}, \\ 
\tilde{p}_{2} & = & y_{\lambda+1}\Upsilon^{\frac{1}{2}}
+(\lambda+1)\Upsilon^{-\frac{1}{2}}, &  
p_{2} & = & -y_{\lambda+1}\tilde{\Upsilon}^{\frac{1}{2}}
-\lambda\tilde{\Upsilon}^{-\frac{1}{2}},   
\end{array}  \label{eq:canotra}  
\end{equation} 
where 
\[ 
\Upsilon= -q_{2}^{2}+8y_{\lambda+1}p_{1} 
+8y_{\lambda+1}^{2}q_{1}-4q_{1}^{2}+2z, \,  
\tilde{\Upsilon}= -\tilde{q}_{2}^{2}-8y_{\lambda+1}\tilde{p}_{1}
+8y_{\lambda+1}^{2}\tilde{q}_{1}-4\tilde{q}_{1}^{2}+2z, 
\]  
and in the left-hand formulae $y_{\lambda+1}$ is to be interpreted as the 
function of $p_{2}$ and $q_{2}$ given by (\ref{eq:yfuncpq}), while  
on the right-hand side take  
\[ 
y_{\lambda+1}=  
\frac{\tilde{p}_{2}\tilde{q}_{2}-(\lambda+1)}{\tilde{q}_{2}^{2}}.  
\]  
For the sake of clarity, we present the (invertible) 
transformations in the  
variables $w_{\lambda}$ and $y_{\lambda}$
diagramatically:
$$ 
\begin{CD}
y_{\lambda} @>{(\ref{eq:mKdVbackl})}>> y_{\lambda +1} \\
@V{(\ref{eq:mmap})}VV      @VV{(\ref{eq:mmap})}V \\
w_{\lambda} @>{(\ref{eq:canotra})}>> w_{\lambda +1}
\end{CD}
$$  
 
It is obvious that (\ref{eq:canotra})   
is a canonical transformation, since the equations for both sets of 
variables are Hamiltonian; it is also possible to verify   
explicitly that the two-form (\ref{eq:twoform}) is preserved.  
A direct calculation also shows that  difference in the  
Hamiltonians is  
\begin{equation}   
h_{\lambda+1}-h_{\lambda}=y_{\lambda+1}, \label{eq:difficha}  
\end{equation}   
which yields (\ref{eq:mKdVytau}) immediately. 
This completes the proof.   
$\square$  

\subsection{The Hamiltonian Systems $\mathcal{H}\mathnormal{_{(i)}}$
and $\mathcal{H}\mathnormal{_{(iii)}}$}   

The Hamiltonian for $\mathcal{H}\mathnormal{_{(i)}}$
is  
\begin{equation}
h_{\lambda}=\frac{1}{2}(p_{1}^{2}+p_{2}^{2})+\frac{1}{2}q_{1}q_{2}^{2}
+\frac{1}{6}q_{1}^{3}
-\frac{\lambda^{2}}{2}q_{2}^{-2}-\frac{1}{2}zq_{1},
\label{eq:hamSK}
\end{equation}
while for $\mathcal{H}\mathnormal{_{(iii)}}$    
it is 
\begin{equation}
H_{\lambda}=\frac{1}{2}(P_{1}^{2}+P_{2}^{2})+\frac{1}{4}Q_{1}Q_{2}^{2}
+\frac{4}{3}Q_{1}^{3}
-2\lambda^{2}Q_{2}^{-2}-zQ_{1}.
\label{eq:hamKK}
\end{equation}
To avoid confusion between the two, we use lower/upper case letters 
for the variables of the systems 
$\mathcal{H}\mathnormal{_{(i)}}$/$\mathcal{H}\mathnormal{_{(iii)}}$ 
respectively.  
It is convenient to use the variables   
\[
w_{\lambda}=q_{1}, \qquad f_{\lambda}=-\frac{1}{2}q_{2}^{2}, \qquad  
W_{\lambda}=Q_{1}, \qquad F_{\lambda}=-\frac{1}{4}Q_{2}^{2}.
\]
Hence, using (\ref{eq:fdef}) 
with the appropriate values of $a$ and $b$,   
\[ 
f_{\lambda}  =  w_{\lambda}''+\frac{1}{2}w_{\lambda}^{2}-\frac{z}{2}, 
\qquad F_{\lambda}  =  W_{\lambda}''+4W_{\lambda}^{2}-z.  
\] 
 
For case (i) the Miura map and its inverse are given by      
\begin{equation}     
w_{\lambda}=-y_{\lambda}'-y_{\lambda}^{2},   
\qquad y_{\lambda}=\frac{f_{\lambda}'+\lambda}{2f_{\lambda}}, 
\label{eq:sap}   
\end{equation}       
where $y_{\lambda}$ is a solution to (\ref{eq:mSK}) 
for $\alpha=\lambda-\frac{1}{2}$. Although case (iii) is related to 
the same modified equation, it will be helpful for deriving the 
B\"{a}cklund transformation to return to the    
original formalism of Section 2, 
where the case (iii)  (with $a=1/4$)   
the Miura map and its inverse are given by  
\begin{equation}   
W_{\lambda}=-Y_{\lambda}'-\frac{1}{2}Y_{\lambda}^{2}, \qquad     
Y_{\lambda}=\frac{F_{\lambda}'+\lambda}{F_{\lambda}}, \label{eq:tap}  
\end{equation}   
where $Y_{\lambda}$ is a solution to (\ref{eq:mKK}) 
for $\alpha=\lambda-1$. The derivation of the B\"{a}cklund 
transformation for the equation (\ref{eq:mSK}) (or equivalently 
(\ref{eq:mKK})) is most easily achieved with the tau-functions, naturally 
related to the tau-functions of the SK/KK hierarchies (see  
\cite{rog} for definitions of these).    
 
\begin{defn}
 For a solution $(q_{j}(z),p_{j}(z))$
of the system $\mathcal{H}\mathnormal{_{(i)}}$,  
with the Hamiltonian $h_{\lambda}(z)=h_{\lambda}(q_{j}(z),p_{j}(z),z)$,
the associated tau-function is given by
\[
h_{\lambda}(z)=-3\frac{d}{dz}\log[\tau_{\lambda}(z)].
\]
\end{defn}

\begin{defn}
 For a solution $(Q_{j}(z),P_{j}(z))$
of the system $\mathcal{H}\mathnormal{_{(iii)}}$, with  
the Hamiltonian $H_{\lambda}(z)=H_{\lambda}(Q_{j}(z),P_{j}(z),z)$,
the associated tau-function is given by
\[
H_{\lambda}(z)=-\frac{3}{2}\frac{d}{dz}\log[\tilde{\tau}_{\lambda}(z)].
\]
\end{defn}

Now we may show: 

\begin{propn}
 The B\"{a}cklund transformation for the equation
(\ref{eq:mSK}) may be written in the form
\begin{equation}
y_{\lambda+3}=y_{\lambda}-\frac{\lambda}{f_{\lambda}}
+\frac{2(\lambda+\frac{3}{2})}{F_{\lambda+\frac{3}{2}}}.
\label{eq:mSKbackl}
\end{equation}
This is related to a canonical transformation between  the
system $\mathcal{H}\mathnormal{_{(i)}}$ with parameter $\lambda$
and the system $\mathcal{H}\mathnormal{_{(iii)}}$ with parameter 
$\lambda-\frac{3}{2}$:
\[
q_{j}\rightarrow Q_{j}, \quad
p_{j}\rightarrow P_{j}, \quad
z\rightarrow z,
\quad
h_{\lambda}\rightarrow H_{\lambda-\frac{3}{2}}.
\]
The modified variable $y_{\lambda}$ may also be written in terms of
the two tau-functions related by this transformation:
\begin{equation}
y_{\lambda}=(\log[\tilde{\tau}_{\lambda-\frac{3}{2}}/\tau_{\lambda}^{2}])'.
\label{eq:mSKytau}
\end{equation}
\end{propn}

\noindent  
{\bf Proof.}  
As for case (ii), $w_{\lambda}$ (and also $W_{\lambda}$) 
may be related to two
different modified variables by the Miura map. Hence we see that  
\begin{eqnarray}        
y_{\lambda}-y_{-\lambda} & = & \frac{\lambda}{f_{\lambda}}, \label{eq:fyrela} 
\\ 
Y_{\lambda}-Y_{-\lambda} & = & \frac{2\lambda}{F_{\lambda}}. \label{eq:fyrelb} 
\end{eqnarray}    
Clearly (\ref{eq:fyrela}) constitutes a B\"{a}cklund transformation for 
(\ref{eq:mSK}), as it relates two solutions for different 
parameter values. However, it is not very 
useful because it cannot be iterated to obtain a sequence 
of solutions. To achieve this requires a canonical tranformation from 
$\mathcal{H}\mathnormal{_{(i)}}$ to $\mathcal{H}\mathnormal{_{(iii)}}$, 
and 
then from $\mathcal{H}\mathnormal{_{(iii)}}$ back 
to $\mathcal{H}\mathnormal{_{(i)}}$ with the 
overall change $\lambda\rightarrow\lambda + 3$.   
First of all observe that on comparing the 
parameters $\alpha$ in (\ref{eq:mSK}) and (\ref{eq:mKK}), it is 
apparent that we may identify  
\[ 
y_{\lambda}=-Y_{-\lambda+\frac{3}{2}}, 
\] 
and so the Miura map for case (iii) implies 
\[ 
W_{-\lambda+\frac{3}{2}}=y_{\lambda}'-\frac{1}{2}y_{\lambda}^{2}. 
\] 

Since $W_{-\lambda+\frac{3}{2}}$  
may be found from $w_{\lambda}$,   
via the formula  
\begin{equation}
W_{-\lambda+\frac{3}{2}}=-w_{\lambda}-\frac{3}{2}y_{\lambda}^{2},
\label{eq:slap}
\end{equation}
it is obvious that     
there is an induced canonical transformation from the system 
with Hamiltonian $h_{\lambda}$ to the system with Hamiltonian 
$H_{-\lambda+\frac{3}{2}}=H_{\lambda-\frac{3}{2}}$. This 
transformation and its inverse may be calculated  
explicitly in terms of coordinates, i.e.    
\[     
\begin{array}{llllll}   
Q_{1} & = & -q_{1}-\frac{3}{2}y_{\lambda}^{2}, &   
q_{1} & = & -Q_{1}-\frac{3}{2}y_{\lambda}^{2}, \\
P_{1} & = & -p_{1}+3y_{\lambda}(q_{1}+y_{\lambda}^{2}),
& 
p_{1} & = & -P_{1}-3y_{\lambda}(Q_{1}+\frac{1}{2}y_{\lambda}^{2}), \\
Q_{2} & = & \aleph^{\frac{1}{2}}, &  
q_{2} & = & \tilde{\aleph}^{\frac{1}{2}},  \\
P_{2} & = & -\frac{1}{2}y_{\lambda}\aleph^{\frac{1}{2}}
-2(\lambda-\frac{3}{2})\aleph^{-\frac{1}{2}}, &  
p_{2} & = & y_{\lambda}\tilde{\aleph}^{\frac{1}{2}}
+\lambda\tilde{\aleph}^{-\frac{1}{2}}, 
\end{array}  
\]  
where
\[
\aleph= -2q_{2}^{2}-12y_{\lambda}p_{1}
-6q_{1}^{2}+6z, \,  
\tilde{\aleph}= -\frac{1}{2}Q_{2}^{2}+6y_{\lambda}P_{1}
-3Q_{1}^{2}+9y_{\lambda}^{2}Q_{1}+\frac{9}{4}y_{\lambda}^{4}+3z,
\]
and in the above $y_{\lambda}$ is to be interpreted alternately  
as a function 
of $p_{2}$, $q_{2}$ and $P_{2}$, $Q_{2}$: 
\[ 
y_{\lambda}=\frac{p_{2}q_{2}-\lambda}{q_{2}^{2}} 
=-\frac{2P_{2}Q_{2}+4(\lambda-\frac{3}{2})}{Q_{2}^{2}} .
\] 
This is the analogue of the transformation 
between autonomous H\'{e}non-Heiles systems considered in   
\cite{baker2,eno}.   
 
The tau-function formula (\ref{eq:mSKytau}) follows directly from 
a calculation of the difference of the two Hamiltonians: 
\[ 
h_{\lambda}-H_{\lambda-\frac{3}{2}}=\frac{3}{2}y_{\lambda}. 
\]   
Then in terms of tau-functions, we have 
\[ 
y_{\lambda+3}-y_{\lambda}=\frac{d}{dz}\left( 
\log[\tilde{\tau}_{\lambda+\frac{3}{2}}/
\tilde{\tau}_{\lambda-\frac{3}{2}}]+2\log[ 
\tau_{\lambda}/\tau_{\lambda+3}] \right). 
\] 
On using the formulae (\ref{eq:fyrela},\ref{eq:fyrelb}), the B\"{a}cklund 
transformation (\ref{eq:mSKbackl}) follows. Thus 
overall there is an induced 
canonical transformation from $\mathcal{H}\mathnormal{_{(i)}}$ to itself, 
with the parameter $\lambda\rightarrow\lambda+3$, as  
can be seen from a diagram:  
$$
\begin{CD}
y_{\lambda} @>{(\ref{eq:fyrela})}>> y_{-\lambda} @=
-Y_{\lambda+\frac{3}{2}} @>{(\ref{eq:fyrelb})}>>
Y_{-\lambda-\frac{3}{2}} @=
-y_{\lambda+3} \\
@V{(\ref{eq:sap})}VV  @V{(\ref{eq:sap})}VV @V{(\ref{eq:tap})}VV
@V{(\ref{eq:tap})}VV @V{(\ref{eq:sap})}VV \\
w_{\lambda} @= w_{-\lambda} @>>{(\ref{eq:slap})}>
W_{\lambda+\frac{3}{2}} @= W_{-\lambda-\frac{3}{2}}
@>>{(\ref{eq:slap})}> w_{\lambda+3}
\end{CD}
$$
Note that in the 
right hand side of (\ref{eq:mSKbackl}), $f_{\lambda}$ may be 
determined entirely in terms of $y_{\lambda}$ and its derivatives, and 
similarly for $F_{\lambda+\frac{3}{2}}$.     
$\square$

\subsection{Analogues of the 
Toda Lattice for Sequences of Tau-Functions} 

It is well known \cite{kaj,oka}
that for PII, the tau-functions related by the
B\"{a}cklund transformation (after rescaling) satisfy the Toda lattice
equation,
\begin{eqnarray}  
D_{z}^{2}\tau_{\lambda}\cdot\tau_{\lambda} 
& = & \tau_{\lambda-1}\tau_{\lambda+1}.
\label{eq:PIItodal} 
\end{eqnarray} 
The tau-functions of (\ref{eq:mKdV}) (related by the B\"{a}cklund  
transformation) also satisfy a 
bilinear lattice equation, which is  
a fourth order analogue of (\ref{eq:PIItodal}). 
Indeed, on the one hand  
(substituting for $w_{\lambda}$ in terms of $\tau_{\lambda}$  
into (\ref{eq:kfdef})) we have   
\begin{equation} 
f_{\lambda}=\tau_{\lambda}^{-2}(D_{z}^{4}\tau_{\lambda}\cdot\tau_{\lambda}) 
-\frac{z}{2}, \label{eq:taukf}  
\end{equation}    
while on the other hand (from (\ref{eq:mKdVytau}) and (\ref{eq:taupm}))    
\begin{equation}    
(\log[\tau_{\lambda+1} 
\tau_{\lambda-1}/\tau_{\lambda}^{2}])'  
=y_{\lambda}+y_{-\lambda}= (\log[f_{\lambda}])'.    
\label{eq:taueffs}  
\end{equation}   
Thus integrating left- and right-hand sides of (\ref{eq:taueffs})  
and comparing with  (\ref{eq:taukf})  
yields the bilinear equation    
\begin{eqnarray}  
2D_{z}^{4}\tau_{\lambda}\cdot\tau_{\lambda}
-z\tau_{\lambda}^{2} & = & k_{\lambda}\tau_{\lambda-1}\tau_{\lambda+1}.
\label{eq:kdvlat} 
\end{eqnarray}  
Note that the constant of integration $k_{\lambda}$ 
may be rescaled arbitrarily,  
since the tau-functions can always be rescaled (without 
affecting $h_{\lambda}$,$y_{\lambda}$). We take the convention 
$k_{\lambda}=1$, which ensures that the tau-functions for the  
rational solutions (presented in section 4)  
are monic polynomials in $z$. However, the   
derivation of (\ref{eq:kdvlat})  
made the generic assumption $f_{\lambda}\not\equiv 0$, 
This may be violated for $\lambda=0$. In this case the constant in 
(\ref{eq:kdvlat}) vanishes (i.e. $k_{0}=0$), and solutions related to PI  
(discussed in the next section) are obtained.       
 
Almost identical arguments lead to the following equations for the 
tau-functions of the systems 
$\mathcal{H}\mathnormal{_{(i)}}$ and 
$\mathcal{H}\mathnormal{_{(iii)}}$:  
\begin{eqnarray}
z\tau_{\lambda}^{4} 
-6\tau_{\lambda}^{2} D_{z}^{4}\tau_{\lambda}\cdot\tau_{\lambda}
+9(D_{z}^{2}\tau_{\lambda}\cdot\tau_{\lambda})^{2}  
& = & \tilde{\tau}_{\lambda-\frac{3}{2}}\tilde{\tau}_{\lambda+\frac{3}{2}},
\label{eq:sklat} \\  
z\tilde{\tau}_{\lambda}^{2} 
-\frac{3}{4}D_{z}^{4}
\tilde{\tau}_{\lambda}\cdot\tilde{\tau}_{\lambda}
& = & \tau_{\lambda-\frac{3}{2}}\tau_{\lambda+\frac{3}{2}} 
\tilde{\tau}_{\lambda}.
\label{eq:kklat}
\end{eqnarray}
We have used the same conventions and genericity assumptions as for  
$\mathcal{H}\mathnormal{_{(ii)}}$.  
Observe that these equations are no longer bilinear, 
but it is consistent  to assign weight one to $\tau$ and weight 
two to $\tilde{\tau}$ in the two equations.

\section{Special Solutions} 

\subsection{Rational Solutions} 
\setcounter{equation}{0}  
   
The rational solutions of the Painlev\'{e} equations have   
been constructed in many different ways  
\cite{abfo,Airault2,kaj}. The rational solutions of   
(\ref{eq:mKdV}) follow a  
pattern very similar to those of PII. They may be generated by  
applying the B\"{a}cklund transformation (\ref{eq:mKdVbackl}) 
repeatedly, starting with the solution $y_{\frac{1}{2}}=0$, to  
produce a sequence of solutions $y_{\lambda}$ for half-integer  
values of $\lambda$. It is  
obvious that the solutions thus generated must be rational, as  
the    
B\"{a}cklund transformation only involves differentiations and  
algebraic operations. The Miura map   
(\ref{eq:mmap}) also yields the associated sequence of  
rational functions  
$w_{\lambda}$ (giving solutions to the system  
$\mathcal{H}\mathnormal{_{(ii)}}$).          
From an algorithmic point of view, it is much more convenient  
to calculate the rational solutions from their   
polynomial tau-functions $\tau_{\lambda}$,  
which can be obtained by iteratively solving 
(\ref{eq:kdvlat}).   
We 
present a few of these rational solutions,  
and their associated tau-functions, in table 1.   
Note that it is sufficient to consider only  
the positive half-integer values of  
$\lambda$, because of the formula   
$y_{\lambda+1}=-y_{-\lambda}$.   

\begin{table}[ht]     
\begin{tabular}{|c|c|c|c|c|c|} \hline
$\lambda$   & $1/2$ &  $3/2$ & $5/2$ & 
$7/2$  & $9/2$ \\ \hline  
 & & & & & \\
$y_{\lambda}$ & 0 & $ -\frac{1}{z}$ & $-\frac{2}{z} $ &
$\frac{-3(z^{5}+96)}{z(z^{5}-144)}$ & 
$ \frac{-4(z^{15}-72z^{10}+217728z^{5}-1741824)} 
{z(z^{15}-1152(z^{10}-84z^{5}+6048))}$     \\
 & & & & & \\ \hline
 & & &  & & \\
$w_{\lambda}$ & 0 & $-\frac{2}{z^{2}}$ & $-\frac{6}{z^{2}}$ &
$\frac{-12(z^{10}+432(z^{5}+8))}{z^{2}(z^{5}-144)^{2}}$ & 
$\frac{-20z^{3}(z^{15}+1008z^{10}+943488z^{5}-47542144)} 
{(z^{10}-1008(z^{5}+48))^{2}}$  \\
 & & & & & \\ \hline
 & & &  &   & \\
$\tau_{\lambda}$ & 1 & $z$ &
$z^{3}$ & $z^{6}-144z$ & $z^{10}-1008(z^{5}+48)$ \\
 & & & & & \\ \hline
\end{tabular}
\caption{Rational solutions for case (ii).} 
\end{table} 

The solutions above can also be obtained  
as specializations of the well-known sequence of rational solutions  
of the mKdV/KdV hierarchy constructed in \cite{admo},  
and thus it is possible to express the polynomial tau-functions 
$\tau_{\lambda}$ as Wronskian determinants \cite{honeth}.      
              
Rational solutions $y_{\lambda}$ of the equation  
(\ref{eq:mSK}), together with the Miura-related    
$w_{\lambda}$,  $W_{\lambda-\frac{3}{2}}$ (giving solutions   
to systems $\mathcal{H}\mathnormal{_{(i),(iii)}}$)   
may be generated in a similar fashion, by applying the  
B\"{a}cklund transformation (\ref{eq:mSKbackl}). Some of these   
solutions are given in table 2. There is one sequence     
for the values $\lambda=3j+\frac{1}{2}$ (for integer $j$), 
generated from the trivial solution $y_{\frac{1}{2}}=0$, and   
a second sequence for $\lambda=3j-\frac{1}{2}$ (which is  
related to the first by (\ref{eq:fyrela})).  
\begin{table}[ht]    
\begin{center}  
\begin{tabular}{|c|c|c|c|c|c|c|c|}\hline
$\lambda$ & $-7/2$ & $-5/2$ & $-1/2$ &
 $1/2$ & $5/2$ & $7/2$ &
$11/2$ \\ \hline
  & & &  & & & & \\
$y_{\lambda}$ & $\frac{4}{z}$ & $\frac{3(z^{5}-24)}{z(z^{5}+36)}$ &
$\frac{1}{z}$ & 0 & $-\frac{2}{z}$ & $-\frac{3}{z}$ &
$\frac{-5z^{4}(z^{5}+216)}{(z^{5}+36)(z^{5}-144)}$ \\
 &  & & & & &  & \\ \hline
 & & & & &  &  & \\
$w_{\lambda}$ & $-\frac{12}{z^{2}}$ & $-\frac{6}{z^{2}}$ & 0 & 0 &
$-\frac{6}{z^{2}}$ & $-\frac{12}{z^{2}}$ &
$\frac{-30z^{3}(z^{5}+576)}{(z^{5}-144)^{2}}$ \\
 & & & &  & &  &  \\ \hline
 & & & & & &  &  \\
$W_{\lambda-\frac{3}{2}}$ & $-\frac{12}{z^{2}}$ &
$\frac{-15z^{3}(z^{5}-144)}{2(z^{5}+36)^{2}}$ & $-\frac{3}{2z^{2}}$ &
0 & 0 & $-\frac{3}{2z^{2}}$ & $-\frac{12}{z^{2}}$ \\
 & &  & & & & & \\ \hline
\end{tabular}
\end{center} 
\caption{Rational solutions for cases (i) and (iii).}
\end{table}  
Once again, it is easier to generate solutions using the  
equations    
(\ref{eq:sklat},\ref{eq:kklat}) for the tau-functions.   
We present some of these in table 3 below  
(using two rows for the pair $\tilde{\tau}_{\lambda\pm\frac{3}{2}}$  
related by (\ref{eq:sklat})).      
\begin{table}[ht]     
\begin{center}  
\begin{tabular}{|c|c|c|c|c|}\hline
$\lambda$ &
 $1/2$ & $5/2$ & $7/2$ &
$11/2$ \\ \hline
  & & & & \\
$\tau_{\lambda}$ & $1$ & $z$ & $z^{2}$ & $z^{5}-144$ \\
 & & & & \\ \hline
 & &  &  & \\
$\tilde{\tau}_{\lambda-\frac{3}{2}}$ & $1$ & $1$ & $z$ & $z^{5}+36$ \\
 & & & & \\ \hline
 & &  &  & \\
$\tilde{\tau}_{\lambda+\frac{3}{2}}$ & $z$ & $z^{5}+36$ & $z^{8}$ &
$z^{16}-2^{7}.3^{2}z^{11}+2^{11}.3^{4}.11z^{6}
+2^{14}.3^{6}.11z$ \\
 & & & & \\  \hline
\end{tabular}
\end{center} 
\caption{Polynomial tau-functions for cases (i) and (iii).}
\end{table}

\subsection{Solutions Related to PI} 
   
All three systems have special 
solutions related to (\ref{eq:peeone}),  
the first Painlev\'{e} transcendent (PI). If we  
consider the system $\mathcal{H}\mathnormal{_{(i)}}$ in the case 
$\lambda=0$, the same substitution that works in the
ordinary (autonomous) system causes the equations of motion to separate.
Putting
\[
Q_{\pm}=q_{1}\pm q_{2}
\]
into Hamilton's equations for $h_{0}$, we find
\[
Q_{\pm}''=-\frac{1}{2}Q_{\pm}^{2}+\frac{z}{2},
\]
which (up to a scaling) is just two separate copies of PI.  
The corresponding solution to (\ref{eq:mSK}) is
\[
y_{0}=(\log[Q_{+}-Q_{-}])',
\]
where we assume that $Q_{+}$ and $Q_{-}$ are not equal.   
So by applying  
the B\"{a}cklund transformation   
(\ref{eq:mSKbackl}) we get the general solution 
to the system 
$\mathcal{H}\mathnormal{_{(i)}}$ for $\lambda=3j$, and to 
$\mathcal{H}\mathnormal{_{(iii)}}$ for $\lambda=3(j+\frac{1}{2})$, for 
all integers $j$. However, there is also the degenerate 
case $Q_{+}=Q_{-}$, for which 
\[ 
f_{0}\equiv 0. 
\] 
This implies that the inverse Miura map and 
the B\"{a}cklund transformation both break down. However, it is still 
possible to obtain a sequence of special solutions for the same 
parameter values, and they are also related to PI. Similarly, starting 
from a degenerate solution $Y_{0}$, corresponding to 
\[ 
F_{0}\equiv 0, 
\] 
the B\"{a}cklund transformation (\ref{eq:mSKbackl}) 
gives a sequence of special solutions 
to the system 
$\mathcal{H}\mathnormal{_{(i)}}$ 
for $\lambda=3(j+\frac{1}{2})$, and to 
$\mathcal{H}\mathnormal{_{(iii)}}$ for $\lambda=3j$. We explain these 
degenerate solutions in more detail for case (ii). 

The degenerate case for all three systems is $\lambda=0$, 
$f_{0}\equiv 0$. So in case (ii) $w=w_{0}$ must satisfy 
\[ 
w''+3w^{2}-\frac{z}{2}=0, 
\] 
which is equivalent to PI (after rescaling $w$ and $z$). The inverse 
Miura map breaks down in this case. However, the ordinary 
Miura map means that $y=y_{0}$  
satisfies the Riccati equation 
\[ 
y'+y^{2}+w_{0}=0. 
\] 
This is linearized by setting $y=(\log[\tau])'$, to yield   
\[ 
\tau''+w_{0}\tau=0, 
\] 
which (for each $w_{0}$) gives  a one parameter family of    
solutions $y_{0}$ to (\ref{eq:mKdV}).   
Because 
$f_{0}\equiv 0$, the B\"{a}cklund 
transformation (\ref{eq:mKdVbackl}) 
breaks down for the solution $y_{0}$. However, we still 
have 
\[ 
y_{\lambda+1}=-y_{-\lambda}, 
\] 
and so we can safely apply the B\"{a}cklund transformation to $y_{1}=-y_{0}$ 
to obtain a sequence of solutions for all integers $\lambda$, as 
well as corresponding solutions to the system 
$\mathcal{H}\mathnormal{_{(ii)}}$.    

\vspace{.1in} 
{\bf Remark.} These   
degenerate solutions to  
(\ref{eq:mKdV}) are the analogues of the Airy function   
solutions to PII (see  \cite{kaj} and some of the other references).  
The fact that solutions of PI give special solutions  
of (\ref{eq:mKdV}) is a particular case of a result in \cite{kudry}.

\section{Conclusions} 

We have found that the Hamiltonian formalism for stationary  
flows of certain integrable PDEs may be extended  
to the similarity solutions.  For the particular examples we have   
considered, non-autonomous generalizations of integrable H\'{e}non-Heiles  
systems are obtained.   
Applying the same technique to the travelling wave similarity solutions  
of (\ref{eq:fif}) leads to H\'{e}non-Heiles Hamiltonians with   
harmonic terms, while the scaling similarity solutions  
of Hirota-Satsuma and a related system correspond to  
non-autonomous versions of the integrable  
Hamiltonians with quartic potentials     
considered in \cite{baker,baker2}; these examples appear  
in \cite{honeth}.  
Our method should always work  
provided that the ODEs for the similarity solutions can be written   
in the form  
\[  
B_{w}\,f=0,   
\] 
where $B_{w}$ is the scaled version of   
the Hamiltonian operator for the PDE, and $f$  
is the gradient of the (scaled) Hamiltonian for the  
PDE plus some extra (possibly non-autonomous) terms (cf. section 2).                 
  
It seems to be a common phenomenon that   
equations of Painlev\'{e} type can be viewed   
as non-autonomous generalizations   
of autonomous, integrable   
Hamiltonian systems. For instance, the   
equations in \cite{harnad} are non-autonomous  
versions of the integrable Neumann systems.   
This connection might be used to study the   
asymptotics of higher order Painlev\'{e}  
equations.   
Recently we have succeeded in extending our    
approach to the multi-phase similarity   
solutions of the kind considered   
by Newell \cite{newell}, but this will be considered (together with  
the connection to isomonodromic deformations) in a   
forthcoming article \cite{hone3}.

\section{Acknowledgements}
This work was supported by an EPSRC studentship.   
It is a pleasure to thank my supervisor, Harry Braden, for all  
his help and encouragement. I would also like to thank   
Allan Fordy for many useful discussions.


\begin{thebibliography}{99}

\bibitem{ars}M.J.~Ablowitz, A.~Ramani and H.~Segur,
 {\it Nonlinear evolution equations and ordinary differential
 equations of Painlev\'{e} type},
 Lett. Nuovo Cim. 23, 333-338 (1978).

\bibitem{abfo}M.J.~Ablowitz and A.S.~Fokas,
 {\it On a unified approach to transformations and
 elementary solutions of Painlev\'{e} equations},
 J. Math. Phys. 23 (1982).

\bibitem{abcla}M.J.~Ablowitz and P.A.~Clarkson,
 {\it Solitons, Nonlinear Evolution Equations and
 Inverse Scattering}, Cambridge University Press (1991).

\bibitem{admo}M.~Adler and J.~Moser,
 {\it On a Class of Polynomials connected with the Korteweg-deVries
 Equation},
 Commun. Math. Phys. 61, 1-30 (1978).

\bibitem{Airault2}H.~Airault,
 {\it Rational Solutions of Painlev\'{e} Equations},
 Stud. Appl. Math. 61, 31-53 (1979).

\bibitem{babe}O.~Babelon and D.~Bernard,
 {\it From form factors to correlation functions. The Ising model},
 Phys. Lett. B288, 113-120 (1992).

\bibitem{baker}S.~Baker, V.Z.~Enolskii and A.P.~Fordy,
 {\it Integrable Quartic Potentials and Coupled KdV Equations},
 Phys. Lett. A201, 167-174 (1995).

\bibitem{baker2}S.~Baker,
 {\it Squared Eigenfunction Representations of Integrable
 Hierarchies}, PhD Thesis, University of Leeds (1995).

\bibitem{caboz}R.~Caboz, L.~Gavrillov and V.~Ravoson,
 {\it Separability and Lax pairs for
 H\'{e}non-Heiles system}, J. Math. Phys. 34, 2385-2393 (1993).

\bibitem{cha}Y.F.~Chang, M.~Tabor and J.~Weiss,
 {\it Analytic structure of the H\'{e}non-Heiles
 Hamiltonian in integrable and nonintegrable regimes},
 J. Math. Phys. 23, 531-538 (1982).

\bibitem{clark}P.A.~Clarkson and M.D.~Kruskal,
 {\it New similarity solutions of the
 Boussinesq equation}, J. Math. Phys. 30, 2201-2213 (1989).

\bibitem{douglas}M.R.~Douglas,
 {\it Strings in Less Than One Dimension and the Generalized KdV
 Hierarchies},
 Phys. Lett. B238, 176-180 (1990).

\bibitem{eno}V.Z.~Enolskii, D.V.~Leykin  and M.~Salerno,
 {\it A canonical transformation between integrable
 H\'{e}non-Heiles systems},
 Phys. Rev. E49, 5897-5899 (1994).

\bibitem{fla}H.~Flaschka and A.C.~Newell,
 {\it Monodromy- and Spectrum-Preserving Deformations I},
 Commun. Math. Phys. 76, 65-116 (1980).

\bibitem{FG}A.P.~Fordy and J.~Gibbons,
 {\it Factorization of operators I},
 J. Math. Phys. 21,
 2508-2510 (1980);
 {\it Factorization of operators II},
 J. Math. Phys. 22, 1170-1175 (1981).

\bibitem{fordy1}A.P.~Fordy,
 {\it The H\'{e}non-Heiles system revisited},
 Physica 52D, 204-210 (1991).

\bibitem{fordy2}A.P.~Fordy,
 {\it Stationary flows: Hamiltonian structures and canonical transformations},
 Physica 87D, 20-31 (1995).

\bibitem{harnad}J.~Harnad, C.A.~Tracy and H.~Widom,
 {\it Hamiltonian Structure of Equations Appearing in Random Matrices},
 NATO ASI Series B, Vol.314, 231-245, Plenum Press (1993).

\bibitem{hone}A.N.W.~Hone,
 {\it Non-autonomous H\'{e}non-Heiles Systems},
 Proceedings of the First Non-Orthodox School of Nonlinearity and
 Geometry, Warsaw (1995).   

\bibitem{honeth}A.N.W.~Hone,
 {\it Integrable Systems and their Finite-Dimensional  
 Reductions}, PhD Thesis, University of Edinburgh (1996).     

\bibitem{hone3}A.N.W.~Hone,   
 {\it Hamiltonians for Multi-Phase Similarity Solutions}, in preparation.  

\bibitem{IIKS}A.R.~Its, A.G.~Izergin, V.E.~Korepin and N.A.~Slavnov,
 {\it Differential Equations for Quantum Correlation Functions},
 Int. J. Mod. Phys. B, Vol. 4, No. 5, 1003-1037 (1990).

\bibitem{kaj}K.~Kajiwara and Y.~Ohta,
 {\it Determinant Structure of the Rational Solutions for the
 Painlev\'{e} II Equation},
 J. Math. Phys. 37, 4693-4704 (1996).

\bibitem{kudry}N.A.~Kudryashov,    
 {\it The first and second Painlev\'{e} equations of higher order   
 and some relations between them},  
 Phys. Lett. A224, 353-360 (1997).   

\bibitem{manin}Y.I.~Manin,
 {\it Sixth Painlev\'{e} Equation, Universal Elliptic Curve and Mirror of
 $P^{2}$},
 {\tt alg-geom/9605010}.

\bibitem{newell}A.C.~Newell,   
 {\it Solitons in Mathematics and Physics},  

\bibitem{oka}K.~Okamoto,
 {\it On the tau-function of the Painlev\'{e} equations},
 Physica 2D, 525-535 (1981);
 {\it Studies on the Painlev\'{e} Equations III}, Math. Ann. 275,
 221-255 (1986).

\bibitem{rog}C.~Rogers and R.~Shadwick,
 {\it B\"{a}cklund transformations and their
 Applications},
 Academic Press (1982).


\end{thebibliography}
\end{document}